\documentclass[preprintnumbers, floatfix, preprintnumbers, letterpaper, superscriptaddress,nofootinbib, twocolumn]{revtex4}
\pdfoutput=1
\usepackage{graphicx}
\usepackage{microtype}
\usepackage{amsmath}
\usepackage{amssymb}
\usepackage{subfigure}
\usepackage{hyperref}
\usepackage{url}
\usepackage{xcolor}
\usepackage{color}
\usepackage{mathrsfs}
\usepackage{calrsfs}
\usepackage{amsfonts}
\usepackage{latexsym}
\usepackage{ragged2e}
\usepackage{epsfig}
\usepackage{textcomp}

\usepackage{caption}
\DeclareCaptionJustification{justified}{\leftskip=0pt \rightskip=0pt \parfillskip=0pt plus 1fil}
\definecolor{vividviolet}{rgb}{0.62, 0.0, 1.0}
\definecolor{amaranth}{rgb}{0.9, 0.17, 0.31}
\definecolor{palatinateblue}{rgb}{0.15, 0.23, 0.89}
\definecolor{brightpink}{rgb}{1.0, 0.0, 0.5}
\definecolor{cornflowerblue}{rgb}{0.39, 0.58, 0.93}
\definecolor{deepcarminepink}{rgb}{0.94, 0.19, 0.22}
\definecolor{radicalred}{rgb}{1.0, 0.21, 0.37}

\hypersetup{ linktoc=all,
    colorlinks, linkcolor={palatinateblue},
    citecolor={brightpink}, urlcolor={amaranth}
}
\newcommand{\changeurlcolor}[1]{\hypersetup{urlcolor=#1}}
\graphicspath{{Images/}}

\renewcommand{\d}[1]{\ensuremath{\operatorname{d}\!{#1}}}

\graphicspath{{Images/}}
\renewcommand{\d}[1]{\ensuremath{\operatorname{d}\!{#1}}}
\makeatletter
\def\@fnsymbol#1{\ensuremath{\ifcase#1\or $\textleaf$ \or \ddagger
\else\@ctrerr\fi}}%

\makeatother

\def\sideremark#1{\ifvmode\leavevmode\fi\vadjust{\vbox to0pt{\vss
 \hbox to 0pt{\hskip\hsize\hskip1em
 \vbox{\hsize1.5cm\tiny\raggedright\pretolerance10000
 \noindent #1\hfill}\hss}\vbox to8pt{\vfil}\vss}}}%
\begin{document}

\title{A Complementary Third Law for Black Hole Thermodynamics}

\author{Yuan \surname{Yao}}
\affiliation{Center for Gravitation and Cosmology, College of Physical Science and Technology, Yangzhou University, \\180 Siwangting Road, Yangzhou City, Jiangsu Province  225002, China}

\author{Meng-Shi \surname{Hou}}
\affiliation{Center for Gravitation and Cosmology, College of Physical Science and Technology, Yangzhou University, \\180 Siwangting Road, Yangzhou City, Jiangsu Province  225002, China}

\author{Yen Chin \surname{Ong}}
\email{ycong@yzu.edu.cn}
\affiliation{Center for Gravitation and Cosmology, College of Physical Science and Technology, Yangzhou University, \\180 Siwangting Road, Yangzhou City, Jiangsu Province  225002, China}
\affiliation{School of Aeronautics and Astronautics, Shanghai Jiao Tong University, Shanghai 200240, China}
\affiliation{Nordita, KTH Royal Institute of Technology \& Stockholm University,
Roslagstullsbacken 23, SE-106 91 Stockholm, Sweden}

\begin{abstract}
There are some examples in the literature, in which despite the fact that
the underlying theory or model does not impose a lower bound on the size of black holes, the final temperature under Hawking evaporation is nevertheless finite and nonzero. 
{We show that under some loose conditions, the black hole is necessarily an effective remnant, in the sense that its evaporation time is infinite.
That is, the final state that there is nonzero finite temperature despite having no black hole remaining cannot be realized. 
We discuss the limitations, subtleties, and the implications of this result, which is reminiscent of the third law of black hole thermodynamics, but with the roles of temperature and size interchanged.
We therefore refer to our result as the ``complemetary third law'' for black hole thermodynamics. }
\end{abstract}

\maketitle

\section{Introduction: The Issue with Temperature of Black Hole Remnants}

In the usual picture of Hawking evaporation, an asymptotically flat Schwarzschild black hole evaporates completely in finite time, although the time scale is extremely long for a stellar mass black hole\footnote{A solar mass non-rotating neutral black hole takes $10^{67}$ years to evaporate, which far exceeds the current age of the Universe $\sim 10^{10}$ years.}. Since the Hawking temperature is inversely proportional to the mass, the black hole becomes hotter as it shrinks. Eventually the energy scale becomes so high that new physics could potentially enter and affect the subsequent evolution. In particular, novel quantum gravity effect may put a stop on Hawking evaporation, thus resulting in a black hole ``remnant''. 

The idea of a black hole remnant can be traced back to the work of Aharonov, Casher and Nussinov \cite{ACN}. It has been suggested that black hole remnants could help to ameliorate the black hole information paradox, though there are arguments against the very existence of remnants. See \cite{1412.8366} for a comprehensive discussion of the debate and a review of various remnant scenarios.

A popular way of obtaining a black hole remnant is via the generalized uncertainty principle (GUP), which incorporates the effect of gravity into the Heisenberg's uncertainty principle. 
Since GUP arises from various general considerations involving gravity and quantum mechanics, as well as string theory \cite{9301067,9305163,9904025,9904026,5,6,7,8,9}, 
it is usually treated as a phenomenological approach to study various properties of quantum gravity.
The simplest form of GUP is given by
\begin{equation}\label{GUP}
\Delta x \Delta p \geqslant \frac{1}{2} \left[\hbar + \frac{\alpha L_p^2 \Delta p^2}{\hbar}\right].
\end{equation}
From here onwards, we set $\hbar=c=G=k_B=1$, unless when explicitly restored for clarity.
Note that if $\alpha \sim O(1)$, as is usually considered in theoretical calculation, then the correction term becomes important at Planck scale. It has been argued that this leads to a correction in the Hawking temperature, resulting in a black hole remnant \cite{pc}. In this scenario however, as the evaporation stops at some finite mass, the temperature also stops at a finite, nonzero value, see Fig.(\ref{masstempalpha}) below. This is somewhat peculiar: a positive temperature seems to suggest that the black hole continues to emit particle, how then does the evaporation completely stop? A possible interpretation is as follows: since the remnant heat capacity vanishes, there is no thermodynamical interaction with its environment. Therefore, it is thermodynamically inert and behaves like an elementary particle \cite{0912.2253}. The finite remnant ``temperature'' should therefore be interpreted as energy of the remnant (via $E=k_BT$). 

If one takes $\alpha < 0$ in Eq.(\ref{GUP}), we would find that there is no lower bound for black hole mass, so in principle the black hole can evaporate completely. However the final temperature is finite and nonzero \cite{1804.05176, 1806.03691}, also see Fig.(\ref{masstempalpha}). 

\begin{figure}[!h]
\centering
\includegraphics[width=3.2in]{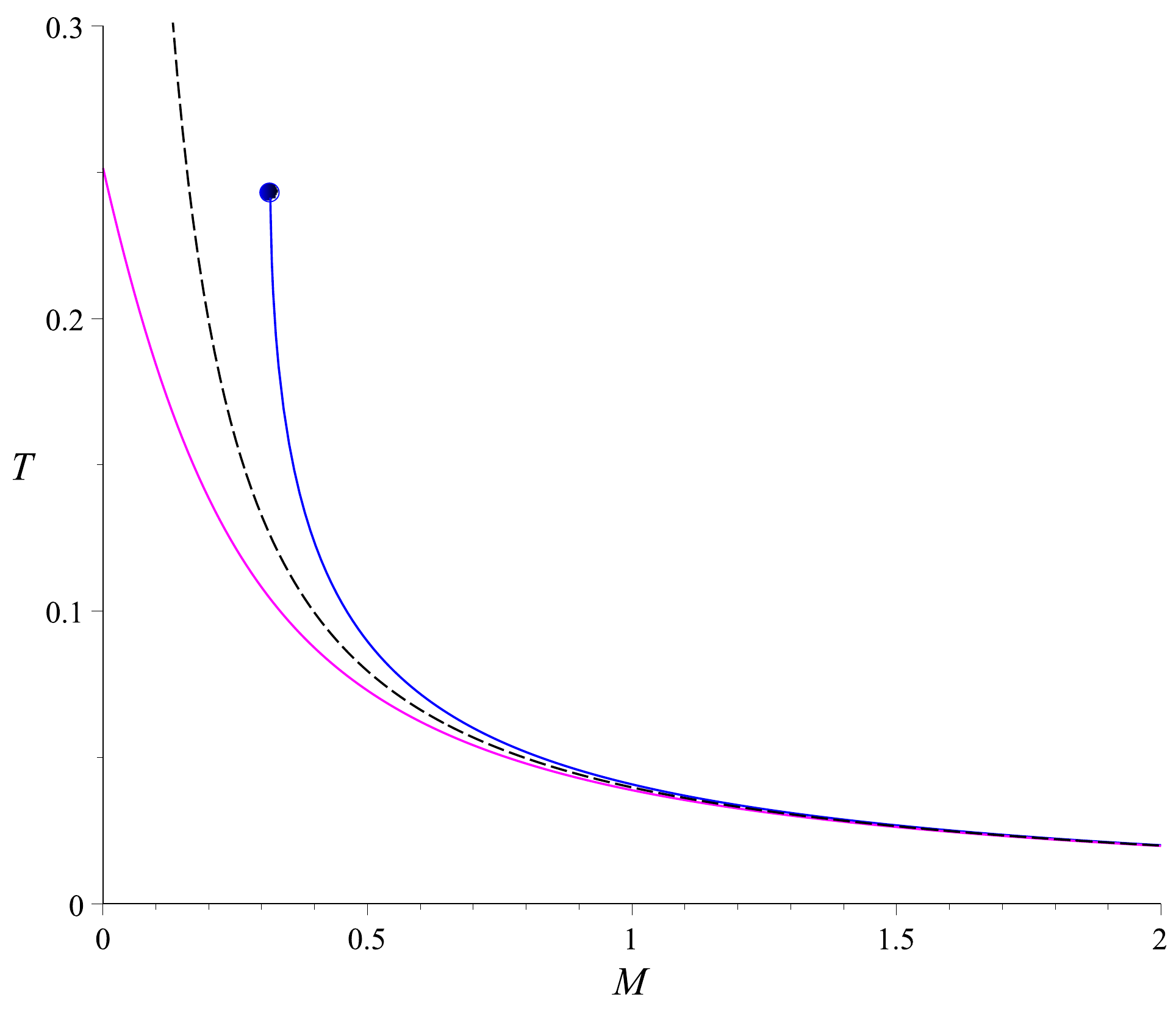}
\caption{The Hawking temperature of an asymptotically flat Schwarzschild black hole. The middle dashed curve corresponds to the usual picture of Hawking evaporation, which diverges as $M \to 0$. The divergence is removed with GUP correction. Specifically, if $\alpha > 0$,  the temperature curve terminates at around $M \sim \sqrt{\alpha}M_p$, as shown by the right-most curve. If $\alpha < 0$, however, GUP correction no longer imposes a lower bound on the black hole size. This corresponds to the left-most curve: the temperature remains finite as the black hole appears to shrink down to zero size.
\label{masstempalpha}}
\end{figure}

Such a choice of sign of $\alpha$ may seem unusual, but it is consistent with some quantum gravity models in which physics at the Planck scale ``classicalized'' and becomes deterministic (as the RHS of the GUP equation goes to zero when $\Delta p \cdot c$ is equal to the Planck energy) \cite{0912.2253, 1804.05176, 1806.03691, 1504.07637}. {Specifically, we have seen that a lattice ``spacetime crystal'' gives rise to such a GUP \cite{0912.2253}. 
Negative GUP parameter is also required if one accepts that Wick-rotation can be applied to obtain GUP-corrected black hole temperature from a Schwarzschild-like black hole (with higher order terms) \cite{1407.0113}.
More recently, it has also been shown that non-commutative geometry \cite{KLVY}, as well as corpuscular gravity, give rise to negative GUP parameter \cite{1903.01382}. 

In addition, even without GUP, the situation that $\Delta x \Delta p \sim 0$ is compatible with other scenarios that have been proposed in the literature, such as the possibility that $\hbar$ is a dynamical field that flows to zero in high energy limit \cite{1212.0454, 1208.5874}; and with Planck mass fixed in 4-dimensions, Asymptotically Safe Gravity with $G \to 0$ is equivalent to $\hbar \to 0$ \cite{0610018} (see also a similar proposal in $f(R)$ gravity \cite{0911.0693}). Incidentally, although some of the very first GUP scenarios came from string theory, in which $\alpha$ is naturally positive, it is not clear that negative $\alpha$ is incompatible with string theory. For example, low energy limit of string theory gives rise to charged dilaton black hole, with string coupling being weak near the singularity. Though we have no right to trust this solution near the singularity, as Horowitz put it \cite{9210119}, it is tempting to speculate whether ``\emph{contrary to the usual picture of large quantum fluctuations and spacetime foam near the singularity, quantum effects might actually be suppressed}.'' This would then be, at least naively, compatible with the $\Delta x \Delta p \sim 0$ scenario at high energy.} 

This can be seen as a virtue of GUP as a phenomenological tool: by taking different signs of  $\alpha$, it can accommodate different kinds of quantum gravity models. The question is: How does one make sense of a nonzero black hole temperature if the black hole has completely evaporated? A possible interpretation is that this is the temperature of the Hawking radiation at the final moment just before the black hole disappears \cite{1804.05176}. This parallels the explanation for the $\alpha>0$ case, but instead of interpreting the temperature as energy of the remnant (since there is no radiation), one now takes it to be the temperature of the final emission of radiation (since there is no black hole). An alternative interpretation as ``vacuum fluctuation'' is discussed further Sec.(\ref{example}).

\begin{figure}[!h]
\centering
\includegraphics[width=3.2in]{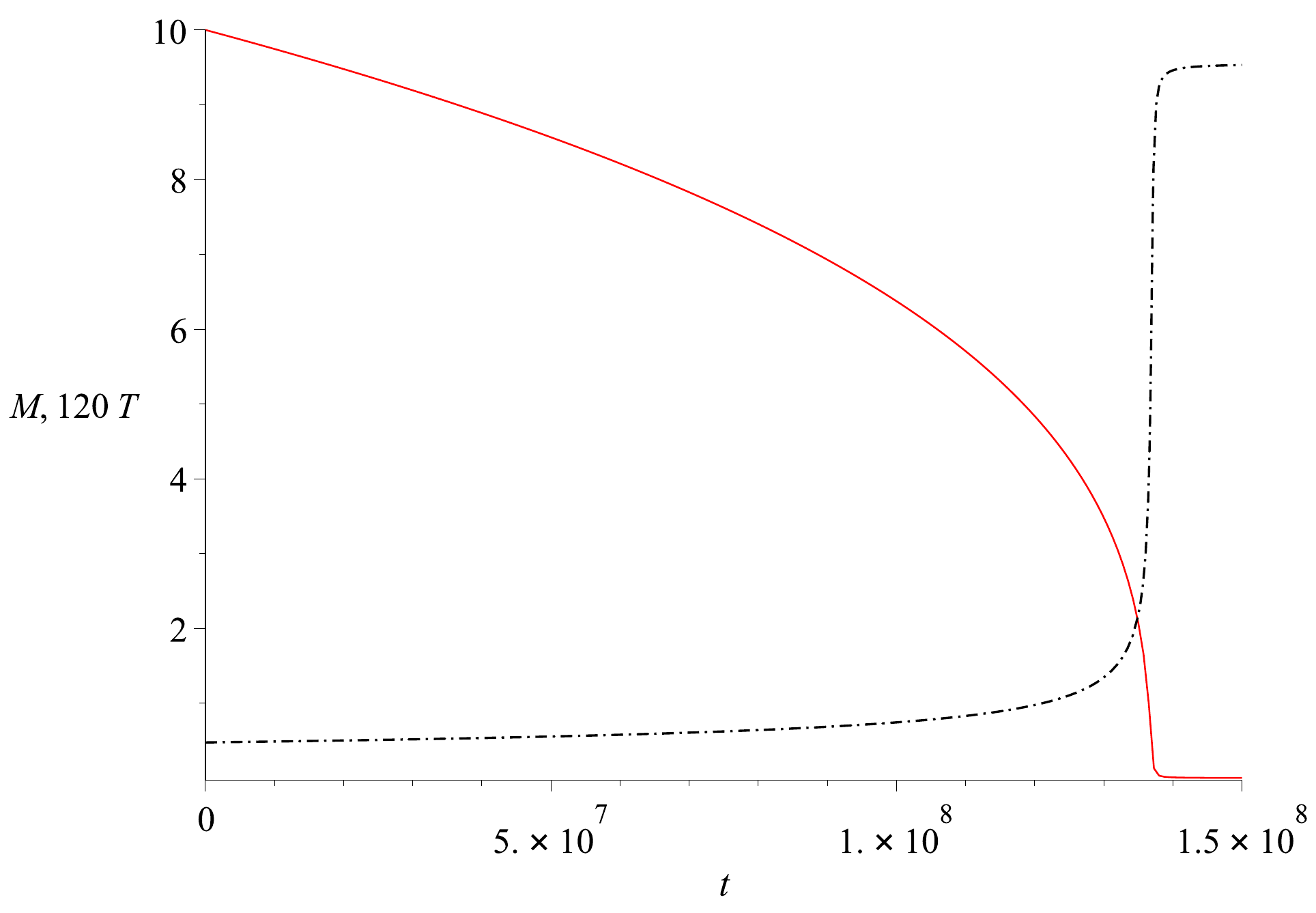}
\caption{The Hawking temperature of an asymptotically flat Schwarzschild black hole with $\alpha=-1$, here in black dash-dotted curve, as a function of time, shows that the temperature tends to a constant value. The mass of the black hole, in red solid curve, tends to zero asymptotically. In order to display both curves in the same diagram, we have multiplied the Hawking temperature by a factor of 120, so that the temperature curve tends to $120 T^*=120/{4\pi} \approx 9.549$.
\label{Mt}}
\end{figure}

However, in \cite{1806.03691}, a more detailed study of the evaporation process reveals that the $\alpha < 0$ GUP-corrected black hole actually takes an infinite amount of time to evaporate completely, so there is no need to resort to the aforementioned interpretation; the black hole simply continues to evaporate indefinitely (indeed its heat capacity is always negative and so it interacts thermodynamically with the environment), with temperature asymptotes to a finite nonzero value. This value is $T^*=1/(4\pi\sqrt{|\alpha|})$ \cite{1804.05176, 1806.03691} (note the typo in \cite{1806.03691}). Thus even though black hole mass is not bounded below, the black hole behaves \emph{effectively} as a meta-stable remnant at late times. See Fig.(\ref{Mt}). This behavior is due to the fact that $\d M/\d t$, though always negative, is not monotonic, and tends to zero as $M \to 0$.

Indeed, the fact that the black hole lifetime is infinite can be shown analytically \cite{1806.03691}. The evolution equation is (a minus sign in missing in \cite{1806.03691}):
\begin{equation}
\frac{\text{d}M}{\text{d}t} = -\frac{M^6}{(4|\alpha| \pi)^4} \left(1-\sqrt{1+\frac{|\alpha| }{M^2}}\right)^4.
\end{equation}
so as $M$ becomes sufficiently small, we have
\begin{equation}
\frac{\text{d}M}{\text{d}t} \sim -\frac{M^2}{(4\pi)^4 {\alpha^2}},
\end{equation}
which leads to
\begin{equation}\label{mgup}
M = M_0\left(\frac{256\pi^4 \alpha^2}{256\pi^4 \alpha^2 + M_0t}\right),
\end{equation}
where $M_0$ is the ``initial'' (small) mass. 

{This leads to a natural question: \emph{how generic is this behavior of having an infinite evaporation time when the final temperature is finite and nonzero? Are there any condition required to ensure this?} Black holes are known to behave like thermodynamical systems, in particular the third law states that a black hole (of course of nonzero mass) cannot reach zero temperature state in finite number of steps. Here we are claiming a complementary result: a black hole cannot reach a state with nonzero temperature but zero mass. This is the \emph{complementary third law} of black hole thermodynamics.}

We found that this behavior is in fact rather general, and can be stated as
\begin{quote}\label{theorem}
\textbf{Theorem:} \emph{Consider an $n$-dimensional neutral static black hole spacetime, with areal radius $r$, and horizon at $r=r_h$.
Assume that the Hawking temperature $T$ and the black hole mass $M$ are analytic functions of $r_h$. 
Suppose $\d M/\d t = -C AT^n$, where $C > 0$ is a constant, and $T \to T^* \in (0, \infty)$ as $r_h \to 0$, then $r_h \to 0$ only if $t \to \infty$, provided that the $k$-th derivative $M^{(k)}$, for $k < n-1$, do not all vanish when $r_h = 0$. }
\end{quote}
We have assumed that there is no problem with convergence of the series expansion. In particular, since $T$ and $M$ are defined only in the domain $[0, \infty)$, all the associated limits and differentiability at 0 are to be understood as being one-sided ($r_h \to 0^+$). Note that Hawking temperature of the usual kind $T \propto 1/M$ is not differentiable at $r_h=0$.

This result reminded us of the (``Nernst version'' of) third law of black hole thermodynamics: zero temperature (extremal) black hole, which is of nonzero size, is unattainable in finite number of steps. Here we have the opposite scenario, zero mass/size\footnote{Zero mass and zero size are not always interchangeable, see Sec.(\ref{discussion}).} black hole is unattainable in finite time\footnote{A finite time is equivalent to a finite number of steps, with each Hawking particle emission counted as one ``step''.} under Hawking evaporation if the temperature is nonzero. We therefore refer to this theorem as a ``complementary third law''. See Sec.(\ref{discussion}) for more discussions.

Here we assume that the Stefan-Boltzmann law for arbitrary spacetime dimension $n \geqslant 4$ holds during the entire evolution \cite{LV, 0510002}, which of course need not be the case; see, e.g., \cite{0004004, 9712017, 1101.1384,0412265}, for a different viewpoint. 
Note that we assumed that the Hawking evaporation is governed by the simple Stefan-Boltzmann law only. This means, for example, we do not study 
 asymptotically de-Sitter spacetimes, in which
Gibbons-Hawking temperature from the cosmological horizon would contribute. Likewise, in an asymptotically locally anti-de Sitter spacetime, the usual reflective boundary condition would complicate the situation (see, however, Sec.(\ref{discussion}) for more discussions).  $A$ denotes the horizon area, with
the constant $C$ incorporating the Boltzmann constant and the greybody factor. The effect of greybody factor, as well as the discreteness of the Hawking radiation (the sparsity \cite{1506.03975, 1512.05809, 1606.01790}) can be ignored since their effects would result in an even longer evaporation time \cite{1806.03691}. In the geometric optic approximation, it is the geometric optic cross section that goes into the Stefan-Boltzmann law, but we also absorb this correction into $C$, as it will not affect the qualitative behavior of the solution. 

In the following, we will first illustrate the theorem with another concrete example, before proving the theorem for the general case. Unlike the GUP-corrected black hole studied in \cite{1804.05176} discussed above, the following example is obtained from classical modified gravity (the Hawking radiation itself is of course semi-classical).

\section{Another Example: A Black Hole Remnant in Massive Gravity}\label{example}

Note that the theorem is quite generic: it does not need the underlying theory to be general relativity. Here for explicitness we show an example in the context of dRGT (de Rham-Gabadadze-Tolley) massive gravity \cite{dRGT0, dRGT, Hassan, 1401.4173}, in which graviton has nonzero mass. 

A dyonic black hole solution in this theory was found in \cite{1610.01505}, with metric coefficient given by
\begin{equation}
-g_{tt}=k + \frac{r^2}{l^2} - \frac{2m_0}{r} + \frac{q_E^2 + q_M^2}{r^2} + m^2\left(\frac{cc_1}{2}r + c^2 c_2\right),
\end{equation}
where $m_0$ is related to the physical mass of the black hole.
Likewise, $q_E$ and $q_M$ are related to the electric and magnetic charges, respectively. In addition, $m$ is essentially the graviton mass, $c$ is a positive constant, while $c_1, c_2$ and $k$ (the sectional curvature of the horizon) can be positive or negative. 
The Hawking temperature for this black hole is \cite{1610.01505}
\begin{equation}
T= \frac{1}{4\pi}\left[\frac{k}{r_+} + \frac{3r_+}{l^2} - \frac{q_E^2+q_M^2}{r_+^3} + m^2\left(cc_1 + \frac{c^2c_2}{r_+}\right)\right].
\end{equation}

{We emphasize that these black holes are ``physical'', in the sense that they have well-behaved thermodynamical properties, as shown in \cite{1610.01505}. Of course, dRGT massive gravity has two metric tensors: a dynamical one, $g_{\mu\nu}$, and a fixed background fiducial metric $f_{\mu\nu}$. The first and foremost requirement (and the entire purpose of dRGT theory) is that the theory should be ghost-free. This is \emph{not} guaranteed for an arbitrary choice of the fiducial metric. In most applications in the context of holography, a degenerate fiducial metric is chosen (as opposed to a Minkowski metric). Such metric was also applied in \cite{1610.01505} to obtain the black hole solution above. Such a choice of $f_{\mu\nu}$ has been demonstrated to also lead to a ghost-free massive gravity theory \cite{1510.03204}. Thus, the theory (and hence the solution) is also physical in the sense that there is no ghost.}

In the absence of magnetic charge, the authors showed that by tuning the various parameters such that
\begin{equation}\label{zero}
m^2c_2 c^2 + k - \Phi_E^2=0,
\end{equation}
where $\Phi_E$ is the electric potential,
one could have a solution with Hawking temperature of the form
\begin{equation}
T = 2r_h P + \frac{m^2 cc_1}{4\pi},
\end{equation} 
where $P=-\Lambda/(8\pi)$ is the pressure term in the extended black hole thermodynamics in an asymptotically locally anti-de Sitter spacetime. In the limit of vanishing horizon $r_h \to 0$, we see that $0<T={m^2 cc_1}/(4\pi) < \infty$. The authors interpreted this as a fluctuation in the temperature of the background spacetime after the black hole has evaporated. That is to say, there is a trace of the black hole left behind if we look at the energy fluctuation of the vacuum. This is of course tiny because graviton mass $m$ is miniscule, although $m$ of order unity was studied in \cite{1610.01505} in the context of holography. (Of course, large AdS black holes with the usual reflective boundary condition will not evaporate in the first place; but see below.) 

The simplest way to achieve Eq.(\ref{zero}) is by considering the neutral case and set $\Phi_E=0$. Take also $\Lambda=0$ so that the pressure term vanishes. We choose $k=1$ (this is enforced by a topological theorem in general relativity once $\Lambda=0$ \cite{Hawking}, but this may not be the case for massive gravity).  Since $c_2$ can be negative, one can set $c_2=-1/(m^2c^2)$. This yields, surprisingly, a \emph{constant} Hawking temperature regardless of the black hole size, namely 
$
T \equiv {m^2 c c_1}/({4\pi}),
$
independent of $r_h$.

To further simplify the calculation, we set the numerical values $m=c=1$, so $c_2=-1$. We remind the readers that our purpose is only to illustrate the aforementioned theorem. It is possible that with this choice of the parameter values the black hole becomes unstable or other problems might arise\footnote{There are indications that dRGT gravity is problematic, since it is plagued with superluminal propagation. In addition, there exist arbitrarily small closed causal curves that result in a lack of well-posed Cauchy problem \cite{1306.5457, 1410.2289, 1505.03518, 1603.03423}.}.  The readers are referred to \cite{1610.01505} for detailed study of the black hole solutions. (Massive gravity also admits a more conventional black hole remnant that tends to zero temperature with finite size \cite{1808.07829}.)

\begin{figure}[!h]
\centering
\includegraphics[width=3.5in]{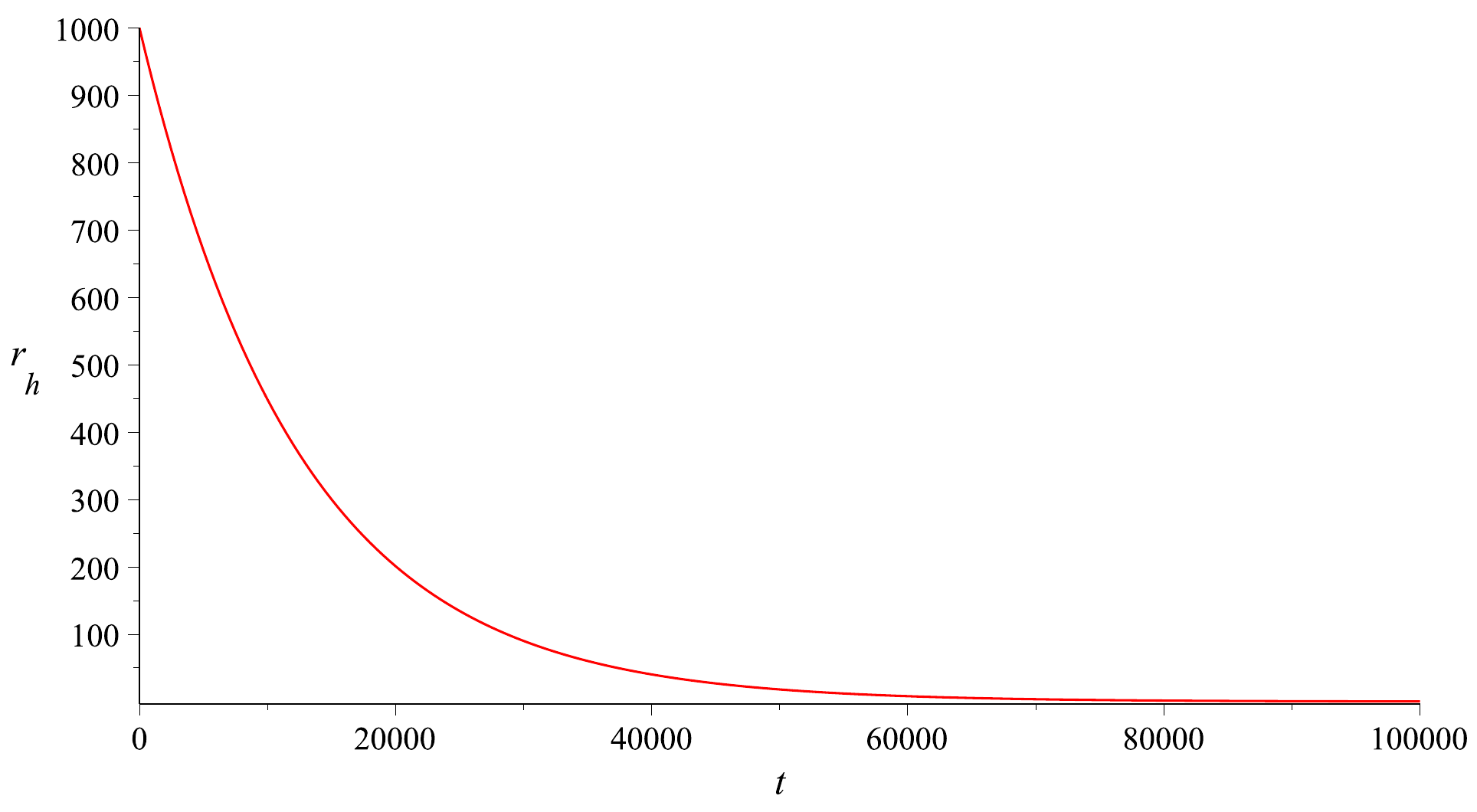}
\caption{The evolution of the massive gravity black hole event horizon radius as function of time. Here we choose $r_h(0)=1000$. 
The black hole parameters are  $m=k=c=1$, $c_2=-1$
The black hole asymptotes to zero size as time goes to infinity. \label{mbh}} 
\end{figure}

The physical mass (the mass that appears in the first law of thermodynamics)  is \cite{1610.01505}
\begin{equation}
M=\frac{r_h}{2}\left[k + \frac{r_h^2}{l^2} + \frac{q_E^2+q_M^2}{r^2} + m^2\left(\frac{cc_1}{2}r_h + c^2 c_2\right)\right],
\end{equation}
which, with our choice of the parameter values, reduces to $M={r_h^2}/{4}$. In Fig.(\ref{mbh}), we set the initial condition $r_h=1000$, and obtain a plot which shows that the horizon tends to zero size only asymptotically.

The analytic proof is straightforward: with $M=r_h^2/4$ and $T=1/(4\pi)$, we have
\begin{equation}
\frac{\text{d}M}{\text{d}t}=\frac{r_h}{2}\frac{\text{d}r_h}{\text{d}t}=-C r_h^2 \frac{1}{(4\pi)^4},
\end{equation}
which yields, with $\tilde{C}=2C/{(4\pi)^4}$,
\begin{equation}
\frac{\text{d}r_h}{\text{d}t} = -\tilde{C} r_h \Longrightarrow \int_{r_h(0)}^\varepsilon \frac{\text{d} r_h}{r_h} = - \tilde{C} \int_{0}^{t^*} \text{d}t,
\end{equation}
where $r_h(0)$ is the initial horizon size, and $t^*$ is the time at which the horizon has shrunk to $\varepsilon$. Integerating yields
\begin{equation}
{\varepsilon} ={r_h(0)} \exp\left[-\tilde{C} t^*\right].
\end{equation}
Therefore, in order to shrink to zero size, $\varepsilon \to 0$, one must have an infinite evaporation time $t^* \to \infty$.

The caveat here is that in this particular example, we have assumed that the area that appears in the Stefan-Boltzmann law is the horizon area. However due to the metric function being 
$-g_{tt}={r}/{2}-{2M}/{r} \sim r/2$ at large $r$, the asymptotic structure is not flat. We know that in AdS, the effective emitting area of black holes with genus $g \geqslant 1$ ($k=-1, 0$) is a constant (essentially the AdS length scale) and independent of the black hole mass \cite{9803061}. Thus, for our example with unusual asymptotic structure, a similar study along the line of \cite{9803061} should be carried out to determine its effective emitting area, which might not simply be the horizon area up to some factors. In other words, our example above may not be correct, but we use it to illustrate how the theorem would work \emph{if} the emission surface is indeed the horizon area.

\begin{figure}[!h]
\centering
\includegraphics[width=3.3in]{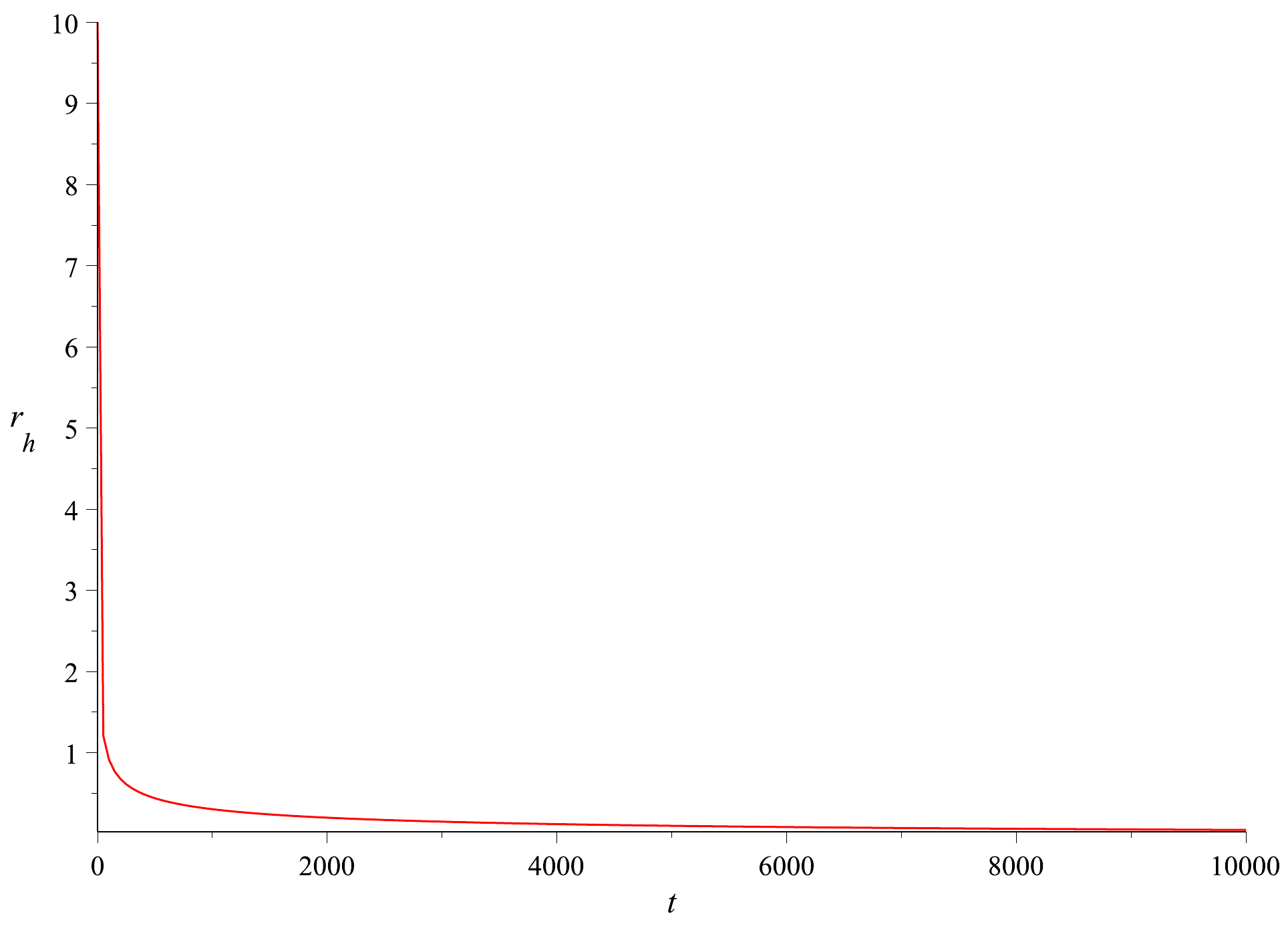}
\caption{The evolution of the black hole event horizon radius as function of time. Here we choose initial radius $r_h(0)=10$. The black hole parameters are
 $m=1=c$,  $c_2=-1$, $l=1$. The black hole also asymptotes to zero size as time goes to infinity. \label{massive2}} 
\end{figure}

We can also choose other values for the various parameters so that the black hole is asymptotically AdS-like, and then modify the boundary conditions to allow the black hole to evaporate (see Sec.(\ref{discussion}) for details). An example with  $m=1=c$,  $c_2=-1$, $k=l=1$ are shown in Fig.(\ref{massive2}). 
Its evolution time is also infinite, but the rate of Hawking evaporation is different from the previous example without the cosmological constant term (the ``pressure'' term). The difference will become clear in the next section after the theorem is proved.

\section{Proof of the Theorem}

We now proceed to prove the theorem. As in the previous example, we will work with $r_h$ in place of $M$, since $M$ is an increasing function of $r_h$. 
(Though this does not imply that $r_h = 0 \Longleftrightarrow M=0$). The dimensionality of spacetime is $n \geqslant 4$.
We assume that $r$ is the areal radius (if not, change to an appropriate coordinate system under which this is true), thus $A \propto r_h^{n-2}$. Absorb the proportional constant into $C$. 
It suffices to consider the late stages of the evolution. Assuming analyticity of the Hawking temperature, we can Taylor expand around $r_h=0$ to obtain
\begin{equation}
T(r_h) = \underbrace{T(0)}_{\neq 0} + T'(0)r_h + \frac{T''(0)}{2} r_h^2 + O(r_h^3), 
\end{equation}
where $T(0)=T^* \in (0, \infty)$ in the statement of the theorem, and prime denotes derivative with respect to $r_h$.
Similarly,
\begin{equation}
M(r_h) = {M(0)} + M'(0)r_h + \frac{M''(0)}{2}r_h^2 + O(r_h^3).
\end{equation}
Taking the derivative yields
\begin{equation}
\frac{\d M(r_h)}{\d t} = \left[M'(0) + M''(0) r_h + O(r_h^2)\right]\frac{\d r_h}{\d t}.
\end{equation}
Then, if $M'(0) \neq 0$, we have
\begin{flalign}\label{drdt}
\frac{\d r_h}{\d t} &= \left[M'(0) + M''(0) r_h + O(r_h^2)\right]^{-1} \notag \\
&~~~~\cdot \left[-C r_h^{n-2} (T^*)^n + O(r_h^{n-1})\right] \notag \\ 
&=-M'(0)^{-1} C r_h^{n-2} (T^*)^n + O(r_h^{n-1}).
\end{flalign}
To lowest order in $r_h$, the differential equation is
\begin{equation}
\frac{\d r_h}{\d t} = -M'(0)^{-1} C r_h^{n-2} (T^*)^n.
\end{equation}
Since $M$ increases with $r$, $M'>0$. In particular, $M'(0) > 0$. So $K:=M'(0)^{-1}C=\text{const}$. Consequently, 
\begin{equation}
\int_{r_h(t_0)}^\varepsilon r_h^{2-n} \d r_h = - K(T^*)^n \int_{t_0}^{t^*} \d t,
\end{equation} 
where $t_0 \gg 1$ is the initial condition at a sufficiently late time where the series approximation is valid. 
Integrating yields
\begin{equation}
\frac{1}{{n-3}}\left({r_h(t_0)^{3-n}}-{\varepsilon^{3-n}}\right) = K(T^*)^n (t_0-t^*).
\end{equation}
It is now clear that if $\varepsilon \to 0$, $t^*$ must tend to infinity (since $n \geqslant 4$). 

In particular, in 4-dimensions, we get,
\begin{equation}
\frac{1}{r_h(t_0)}-\frac{1}{\varepsilon} = K(T^*)^4 (t_0-t^*).
\end{equation}
This is the case for the GUP-corrected black hole in \cite{1806.03691} with negative GUP parameter $\alpha$, since $r_h=2M$, the same as the usual Schwarzschild black hole; c.f. Eq.(\ref{mgup}). 

If $M'(0)=0$, Eq.(\ref{drdt}) would lead to, to lowest order of $r_h$, the differential equation
\begin{equation}
\frac{\d r_h}{\d t} = -M''(0)^{-1} C r_h^{n-3} (T^*)^n.
\end{equation}
This leads to 
\begin{equation}
\int_{r_h(0)}^\varepsilon r_h^{3-n} \text{d} r_h= -\tilde{K}(T^*)^n\int_{t_0}^{t^*} \text{d}t,
\end{equation}
where $\tilde{K}:=M''(0)^{-1} C$.
Integrating yields
\begin{equation}
\frac{1}{{n-4}}\left({r_h(t_0)^{4-n}}-{\varepsilon^{4-n}}\right) = \tilde{K}(T^*)^n (t_0-t^*).
\end{equation}
The evaporation time is clearly infinite for $n \geqslant 5$.
In 4-dimensions, $\varepsilon$ is exponential in $-t^*$. Thus, we again obtain an infinite evaporation time. 
This is the case for the massive gravity black hole example illustrated in Fig.(\ref{mbh}).

However, if both $M'(0)$ and $M''(0)$ vanish, and $M'''(0)\neq 0$, then the integration gives
\begin{equation}
\frac{1}{{n-5}}\left({r_h(t_0)^{5-n}}-{\varepsilon^{5-n}}\right) = \text{const.}(t_0-t^*).
\end{equation}
Note that in 4-dimensions, the corresponding result is
\begin{equation}
\varepsilon - r_h(t_0) = \tilde{K}(T^*)^4(t_0-t^*)
\end{equation}
for some constant $\tilde{K}$. Therefore $t^*$ is now \emph{finite} as $\varepsilon \to 0$:
\begin{equation}
t^*[\varepsilon=0] = \frac{\tilde{K}(T^*)^4t_0 + r_h(t_0)}{\tilde{K}(T^*)^4} < \infty.
\end{equation}
In $n \geqslant 5$ the evaporation time is still infinite.

Indeed, in general, one can see that  as long as the lowest order of nonzero $M^{(k)}(0)$ is $k = {n-1}$, the evaporation time will be finite. This completes the proof.
In particular, this implies that in 4-dimensions, the evaporation is infinite if $M'(0)$ and $M''(0)$ do not both vanish.
{
\section{The Complementary Third Law: Applicability and Subtleties }\label{discussion}}

{To summarize our findings so far:} in this work, we investigated the conditions for a black hole to have ``left-over'' nonzero and finite temperature at the end of Hawking evaporation, at which point the black hole shrinks to zero size. To our knowledge (and that of the authors of \cite{1610.01505}), the massive gravity black hole discussed in Sec.(\ref{example}) is the only known example in classical modified gravity with such a property. The GUP corrected black hole studied in \cite{1804.05176,1806.03691,1504.07637} provided another example from a quantum gravitational correction. {This result parallels the third law of black hole thermodynamics in which (nonzero mass) extremal black hole with zero temperature cannot be attained in finite time, but with the role of mass and temperature reversed, we therefore dubbed it the ``complementary third law of black hole thermodynamics''.}

To be more specific, we found that if the first and the second derivative of $M(r_h)$ do not both vanish at $r_h=0$, then the evaporation time in 4-dimensions is actually infinite (analogously in higher dimensions), and so the black hole behaves as an effective, meta-stable remnant. This result does not assume the underlying theory to be general relativity. It is simply a consequence of the mathematical properties of the Stefan-Boltzman differential equation. Our proof relies on the assumption that $M(r_h)$ and $T(r_h)$ are both real analytic functions, which are infinitely differentiable at $r_h=0$.
Although most calculations in physics literature make use of Taylor expansion almost ubiquitously, this assumption might be too strong. Perhaps one could relax it and the theorem would still remain true. An alternative proof without the use of series expansion would be welcomed so the issues of convergence can be avoided altogether. {Now we shall discuss a few more aspects of the complementary third law. }\newline

\subsection{Charged and Dilaton Black Holes:\\ Nonzero Mass vs. Nonzero Size}

It is worth emphasizing that this phenomenon is somewhat opposite to that of the third law of black hole thermodynamics, in which the black hole size remains finite (and nonzero) and $T=0$ cannot be achieved; whereas here the temperature remains finite (and nonzero) and $r_h=0$ cannot be achieved. The main difference is that our theorem only concerns neutral black holes,
whereas for the third law, it applies to charged and rotating black holes (the only way to get zero temperature in general relativity\footnote{One could have a zero temperature black hole in the zero size limit, if, for example, higher order curvature terms are included in the action \cite{1704.02967}. It is also possible to obtain zero temperature black hole at some nonzero mass without any gauge field in modified gravity theories, e.g., asymptotically safe gravity with higher derivative terms \cite{1007.1317} and in conformal (Weyl) gravity \cite{1811.07309v1} (note that entropy vanishes does not always imply zero area for modified gravity black holes). The usual third law applies to these black holes -- they have infinite lifetime.}).
In the presence of electrical charges and other gauge fields, more analysis would be required to study whether the complementary third law holds, since the evolution under Hawking evaporation would be considerably more complicated. Even for asymptotically flat Reissner-Nordstr\"om black holes, Hiscock and Weems showed that there are charge loss and mass loss regimes, so the ratio $Q/M$ is not necessarily monotonic in time (the evolution is governed by coupled differential equations in certain range of the parameters) \cite{HW}. 

{However, one special charged black hole solution -- the GHS black hole (see below) -- is worth a separate mention, since it gives us the opportunity to point out that the complementary third law is really stating that the final state with zero mass but nonzero temperature cannot be attained, \emph{not} zero size. Usually zero mass also coincides with zero size, but this is not always the case, and therefore the distinction is important.} 

The charged dilatonic ``GHS'' (Garfinkle-Horowitz-Strominger) black hole \cite{ghs1,ghs2,ghs3}, obtained in a low energy limit of string theory, with metric tensor in a Schwarzschild-like coordinate system $\left\{t,r,\theta,\varphi\right\}$:
\begin{flalign}
\d s^2 =& -\left(1-\frac{2M}{r}\right)\d t^2 + \left(1-\frac{2M}{r}\right)^{-1} \d r^2\notag \\ &+ r\left(r-\frac{Q^2}{M}\right)(\d \theta^2 + \sin^2\theta \d\varphi^2).
\end{flalign}
The Hawking temperature is always $T=1/(8\pi M)$ independent of the charge, even in the zero size limit. Note that the horizon stays fixed at $r=2M$, but $r$ is not an areal radius, so one has to transform via $r^2 \mapsto R^2:=r(r-Q^2/M)$ to a coordinate $\left\{t,R,\theta,\varphi\right\}$ in which the areal radius is $R=\sqrt{4M^2-2Q^2}$, which goes to zero in the extremal limit $|Q|=\sqrt{2}M$, with \emph{nonzero} temperature (the temperature is not affected under this change of coordinate). This is very different from extremal Reissner-Nordstr\"om black hole which has zero temperature. 

Despite $\d M/\d t$ being independent of $Q$, charge loss can occur from spontaneous charge particle emission \`a la Schwinger, and as shown by Hiscock and Weems \cite{HW}, this may affect the evolution under Hawking evaporation just like in the Reissner-Nordstr\"om case. Thus our theorem does not strictly apply. 

\begin{figure}[!h]
\centering
\includegraphics[width=3.3in]{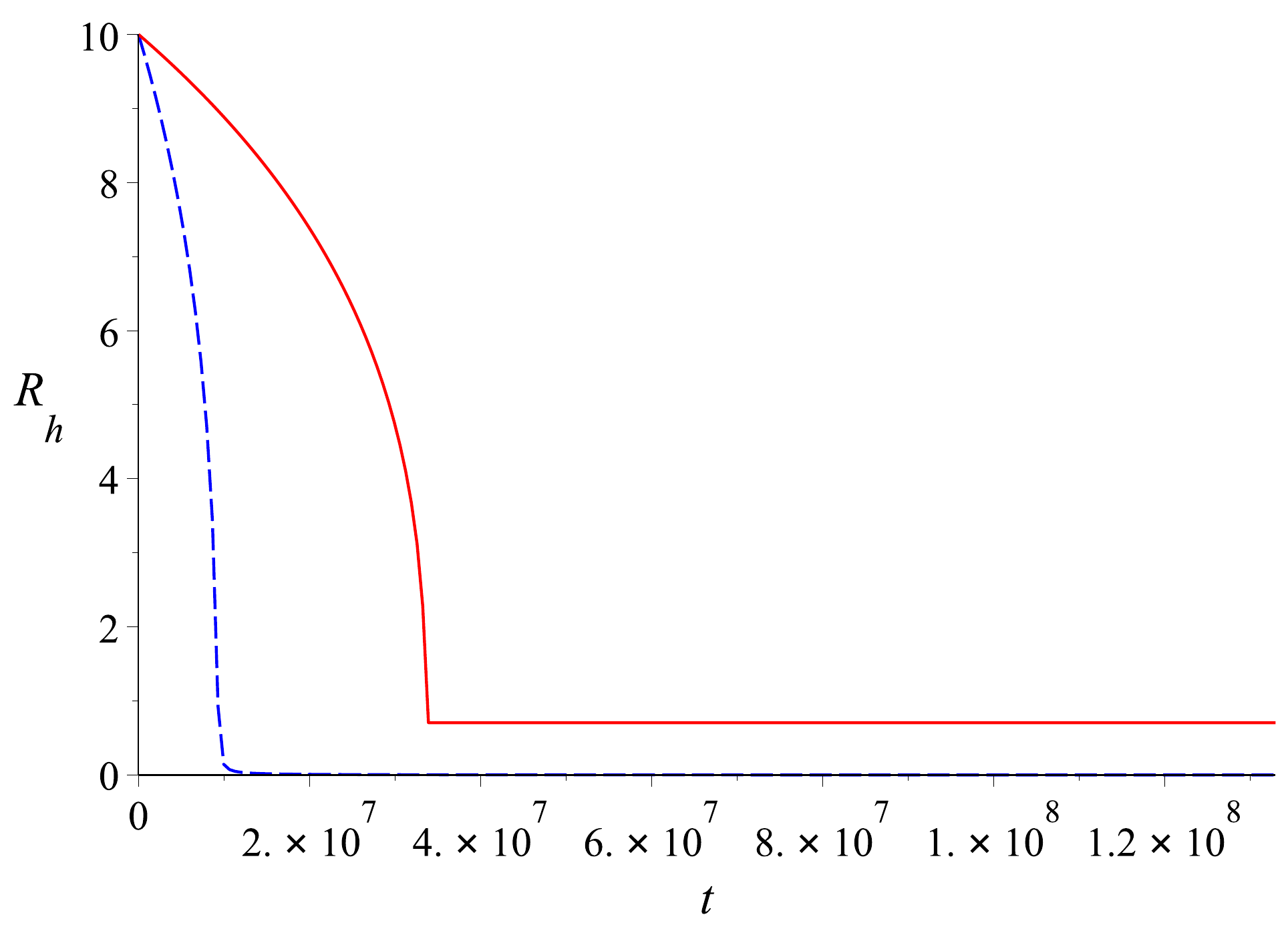}
\caption{The evolution of the areal radius (blue dashed curve) and the mass (red solid curve) of a charged dilatonic GHS black hole as functions of time, assuming that there is no charge loss. Here we choose initial mass to be $M(0)=10$ and $Q$=1. The mass thus asymptotes to its lower bound $M=Q/\sqrt{2}\approx 0.7071$.  The evolution takes an infinite amount of time. \label{ghs}} 
\end{figure}

Nevertheless, if we ignore these subtleties\footnote{However, since $T$ does not depend on $Q$ here, as opposed to the Reissner-Nordstr\"om case, the Schwinger process is not expected to affect the result by much, at least for large enough $M$, in the regime where it is suppressed \cite{Gibbons}. This requires a further investigation beyond the scope of the current work, and will be addressed elsewhere.}, and consider the mass loss of GHS black hole to be governed only by Stefan-Boltzmann law, then our theorem \emph{does} apply. Numerically the results are shown in Fig.(\ref{ghs}). Note the peculiarity that $R \to 0$ but $M$ tends to a finite value, as the black hole approaches a null singularity of zero size, which somehow supports the mass.

\subsection{Non-Examples of Complementary Third Law: \\Anti-de Sitter Black Holes}

{It might be illuminating to also discuss here some \emph{non-examples} that do not satisfy the premise of the complementary third law (note that this is not the same as ``violating'' the law). }
In view of the popularity of holography, asymptotically locally AdS spacetimes are an important class of solutions. 
As we mentioned in the Introduction, we assumed that the evolution is governed by the simple Stefan-Boltzmann law. In asymptotically locally AdS spacetimes, our analysis can be carried over by replacing the usual reflective boundary condition with an absorbing boundary condition, which allows large black holes to evaporate (see, e.g., \cite{0804.0055, 1304.6483, 1307.1796}). In holography, this can be done by coupling
the boundary field theory with an auxiliary system (``AUX'' \cite{1304.6483}), such as another field theory. 

It was previously found that regardless of
their horizon topologies, neutral AdS black holes in such spacetimes take about the same
amount of time to evaporate down to the same size of order $l$, the AdS length scale \cite{1507.02682, 1507.07845}. For positively curved ($k=1$) case, the black hole take about the same amount of time to completely evaporate regardless of its initial mass \cite{1507.02682, 1507.07845}. For flat ($k=0$) case, the evaporation time is infinite. 

Nevertheless these black holes do not satisfy the premise of the theorems. This is because their temperature, in $n$-dimensional spacetime, is given by \cite{9808032}
\begin{equation}
T=\frac{1}{4\pi l^2 r_h}\left[(n-1)r_h^2 + (n-3)kl^2\right],
\end{equation}
which either goes to zero (for $k=0$) or diverges (for $k= 1$) as $r_h \to 0$. (For $k=-1$ case, the black hole tends to a minimum size as the temperature goes to zero.) \cite{1507.07845}
In addition, for $k\in \left\{-1, 0\right\}$, the effective emitting surface area of the black hole is independent of the mass, and is completely fixed by the cosmological constant \cite{9803061}. 
The complementary third law therefore does not apply to AdS black holes, at least the simplest ones addressed here. 

\subsection{Comparison with Conventional Thermodynamics}

It came at a surprise when Bardeen, Carter and Hawking \cite{bch} discovered that black holes satisfy properties that are analogous to thermodynamics. Subsequent discovery that black holes do radiate when quantum mechanics is taken into account, established that black holes \emph{are} thermodynamical system. In particular, black holes satisfy a form of third law:  zero temperature configuration cannot be reached in a finite number of step. This parallels the ``Nernst version'' of third law in conventional thermodynamics. What about the complementary third law?  Is there an analogous phenomenon in other thermodynamical system? 

By definition, in classical thermodynamics, the temperature is related to the entropy by $1/T=(\partial S/\partial U)|_{V,N}$, where $U$ is the internal energy of the system, and the volume $V$, as well as particle number $N$, are held fixed.  It would seem that nonzero temperature (with nonzero internal energy) always corresponds to nonzero entropy. In the context of black holes, nonzero entropy means nonzero area. 

The complementary third law essentially says that black holes with zero entropy yet with nonzero temperature cannot be attained. This is therefore consistent with conventional thermodynamics. 

Note that if we consider quantizing the system, the entropy can be zero for small temperature if the first excited state of the system has energy higher than $k_BT$. This does not concern us since black hole thermodynamics is analogous to classical ordinary thermodynamics.

\section{Discussions: Some Remaining Puzzles}

The question remains for the case $M'(0)=M''(0)$ in 4-dimensions (and analogously in higher dimensions). Black holes that satisfy this property will evaporate in a finite time, and leave behind a nonzero finite temperature. 
What is the correct interpretation for such a temperature? Is it the temperature of the last bit of radiation, or an energy fluctuation of the ambient spacetime (now without a black hole -- so it would be a kind of ``vacuum memory'')? 

Presumably if the final temperature is very low, the second interpretation (to our knowledge, first proposed in \cite{1610.01505}) is plausible, but what if the temperature is ``high'', say $1^\circ \text{C}$ (note that the theorem does not constrain the value of $T^*$ other that it is nonzero and finite), how can this be a ``fluctuation'' of the vacuum? It would be good to have an explicit black hole solution of this type for a detailed study, if it exists.

Let us speculate on a possibility: since black holes contain an enormous volume \cite{1411.2854, 1503.01092} (which does \emph{not} decrease even as the black hole evaporates \cite{9405007v2, 1503.08245, 1604.07222}), the end state of the evolution could be a baby universe that pinches off from the original universe. Perhaps such a pinch-off leaves a finite temperature signature behind in the parent universe. On the other hand, we could turn this around and say that, if the complementary third law is true generically, then maybe there is no 4-dimensional black hole solution that would satisfy  $M'(0)=M''(0)$, and similarly for higher dimensions.

Another issue concerns the singularity of the black hole. As black hole evaporates, does it leave behind a naked singularity? In the usual Schwarzschild case, since temperature becomes extremely high, one may invoke new physics and hope that the even if singularity wasn't already cured by quantum gravity in the first case, will evaporate away together with the horizon. However, if the final temperature remains mild, it leaves open the possibility that naked singularity may form, thus violating the cosmic censorship conjecture. 

To summarize, the complementary third law is more restricted than the standard third law in two ways: firstly, we only study Hawking evaporation, not other physical processes. We restricted our study to the
static case in $n \geqslant 4$ spacetime dimensions, and only to neutral black holes (though the result might hold in more general cases). 
Secondly, if we expresse the black hole mass $M$ as a function of its horizon $M(r_h)$, the complementary third law can be violated if the derivatives $M^{(k)}(0)=0$ for $k < {n-1}$, but then the standard third law may also be violated under certain circumstances \cite{davies, FH, P, Torii}. In a way, the shortcoming of the complementary third law is a virtue since it gives a clear condition for its violation. 
A further study into these conditions and their physical interpretations could yield a deeper understanding into black hole thermodynamics. 

\begin{acknowledgments}
YCO thanks the National Natural Science Foundation of China (grant No.11705162) and the Natural Science Foundation of Jiangsu Province (No.BK20170479) for funding support. 
YCO also thanks Nordita, where the initial phase of this work was carried out, for hospitality during his summer visit while participating at the Lambda Program. He thanks Ingemar Bengtsson for related discussion and Brett McInnes for some suggestions.
The authors thank members of Center for Gravitation and Cosmology (CGC) of Yangzhou University (\href{http://www.cgc-yzu.cn}{http://www.cgc-yzu.cn}) for various discussions and supports.
\end{acknowledgments}


\begin{thebibliography}{99}

\bibitem{ACN}
 Yakir Aharonov, Aharon Casher, Shmuel Nussinov, ``The Unitarity Puzzle and Planck Mass Stable Particles'', {\hypersetup{urlcolor=vividviolet}\href{https://www.sciencedirect.com/science/article/pii/0370269387913207}{Phys. Lett. B \textbf{191} (1987) 51}}.

\bibitem{1412.8366}
Pisin Chen, Yen Chin Ong, Dong-han Yeom, ``Black Hole Remnants and the Information Loss Paradox'', {\hypersetup{urlcolor=vividviolet}\href{https://www.sciencedirect.com/science/article/pii/S0370157315004391?via\%3Dihub}{Phys. Rept. \textbf{603} (2015) 1}}, \href{https://arxiv.org/abs/1412.8366}{[arXiv:1412.8366 [gr-qc]]}.

\bibitem{9301067}
Michele Maggiore, ``A Generalized Uncertainty Principle in Quantum Gravity'',  {\hypersetup{urlcolor=vividviolet}\href{https://www.sciencedirect.com/science/article/pii/0370269393914018?via\%3Dihub}{Phys. Lett. B \textbf{304} (1993) 65}}, \href{https://arxiv.org/abs/hep-th/9301067}{[arXiv:hep-th/9301067]}.

\bibitem{9305163}
Michele Maggiore, ``Quantum Groups, Gravity, and the Generalized Uncertainty Principle'', {\hypersetup{urlcolor=vividviolet}\href{https://journals.aps.org/prd/abstract/10.1103/PhysRevD.49.5182}{Phys. Rev. D \textbf{49} (1994) 5182}}, \href{https://arxiv.org/abs/hep-th/9305163}{[arXiv:hep-th/9305163]}.

\bibitem{9904025}
Fabio Scardigli, ``Generalized Uncertainty Principle in Quantum Gravity from Micro-Black Hole Gedanken Experiment'', {\hypersetup{urlcolor=vividviolet}\href{https://www.sciencedirect.com/science/article/pii/S0370269399001677?via\%3Dihub}{Phys. Lett. B \textbf{452} (1999) 39}}, \href{https://arxiv.org/abs/hep-th/9904025}{[arXiv:hep-th/9904025]}.

\bibitem{9904026} 
Ronald J. Adler, David I. Santiago, ``On Gravity and the Uncertainty Principle'', {\hypersetup{urlcolor=vividviolet}\href{https://www.worldscientific.com/doi/abs/10.1142/S0217732399001462}{Mod. Phys. Lett. A \textbf{14} (1999) 1371}}, \href{https://arxiv.org/abs/gr-qc/9904026}{[arXiv:gr-qc/9904026]}. 

\bibitem{5} Gabriele Veneziano, ``A Stringy Nature Needs Just Two Constants'', {\hypersetup{urlcolor=vividviolet}\href{http://iopscience.iop.org/article/10.1209/0295-5075/2/3/006/meta}{Europhys. Lett. \textbf{2} (1986) 199}}. 

\bibitem{6} David J. Gross, Paul F. Mende, ``String Theory Beyond the Planck Scale'', {\hypersetup{urlcolor=vividviolet}\href{https://www.sciencedirect.com/science/article/pii/0550321388903902}{Nucl. Phys. B \textbf{303} (1988) 407}}.

\bibitem{7} Daniele Amati, Marcello Ciafolini, Gabriele Veneziano, ``Can Spacetime be Probed Below the String Size?'', {\hypersetup{urlcolor=vividviolet}\href{https://www.sciencedirect.com/science/article/pii/037026938991366X}{Phys. Lett. B \textbf{216} (1989) 41}}.

\bibitem{8} Kenichi Konishi, Giampiero Paffuti, Paolo Provero , ``Minimum Physical Length and the Generalized Uncertainty Principle in String Theory'', {\hypersetup{urlcolor=vividviolet}\href{https://www.sciencedirect.com/science/article/pii/0370269390919274}{Phys. Lett. B \textbf{234} (1990) 276}}.

\bibitem{9} Edward Witten, ``Reflections on the Fate of Spacetime'', {\hypersetup{urlcolor=vividviolet}\href{https://physicstoday.scitation.org/doi/10.1063/1.881493}{Phys. Today \textbf{49} (1996) 24}}. 

\bibitem{pc}
Ronald J. Adler, Pisin Chen, David I. Santiago, ``The Generalized Uncertainty Principle and Black Hole Remnants'', {\hypersetup{urlcolor=vividviolet}\href{https://link.springer.com/article/10.1023\%2FA\%3A1015281430411}{Gen. Rel. Grav. \textbf{33} (2001) 2101}}, \href{https://arxiv.org/abs/gr-qc/0106080}{[arXiv:gr-qc/0106080]}.

\bibitem{0912.2253}
Petr Jizba, Hagen Kleinert, Fabio Scardigli, ``Uncertainty Relation on World Crystal and its Applications to Micro Black Holes'', {\hypersetup{urlcolor=vividviolet}\href{https://journals.aps.org/prd/abstract/10.1103/PhysRevD.81.084030}{Phys. Rev. D \textbf{81} (2010) 084030}}, \href{https://arxiv.org/abs/0912.2253v2}{[arXiv:0912.2253 [hep-th]]}.

\bibitem{1804.05176}
Yen Chin Ong, ``Generalized Uncertainty Principle, Black Holes, and White Dwarfs: A Tale of Two Infinities'', {\hypersetup{urlcolor=vividviolet}\href{http://iopscience.iop.org/article/10.1088/1475-7516/2018/09/015/meta}{JCAP \textbf{09} (2018) 015}}, \href{https://arxiv.org/abs/1804.05176}{[arXiv:1804.05176 [gr-qc]]}.

\bibitem{1806.03691}
Yen Chin Ong, ``An Effective Black Hole Remnant via Infinite Evaporation Time due to Generalized Uncertainty Principle'', {\hypersetup{urlcolor=vividviolet}\href{https://link.springer.com/article/10.1007\%2FJHEP10\%282018\%29195}{JHEP \textbf{10} (2018) 195}}, \href{https://arxiv.org/abs/1806.03691}{[arXiv:1806.03691 [gr-qc]]}.

\bibitem{1504.07637}
Bernard J. Carr, Jonas Mureika, Piero Nicolini, ``Sub-Planckian Black Holes and the Generalized Uncertainty Principle'', {\hypersetup{urlcolor=vividviolet}\href{https://link.springer.com/article/10.1007\%2FJHEP07\%282015\%29052}{JHEP \textbf{07} (2015) 052}}, \href{https://arxiv.org/abs/1504.07637}{[arXiv:1504.07637 [gr-qc]]}.

\bibitem{1407.0113}
Fabio Scardigli, Roberto Casadio, “Gravitational tests of the Generalized Uncertainty Principle”, {\hypersetup{urlcolor=vividviolet}\href{https://link.springer.com/article/10.1007\%2FJHEP07\%282015\%29052}{Eur. Phys. J. C \textbf{75} (2015) 425}}, \href{https://arxiv.org/abs/1504.07637}{arXiv:1407.0113 [hep-th]}.

\bibitem{KLVY}
T. Kanazawa, G. Lambiase, G. Vilasi, A. Yoshioka, ``Noncommutative Schwarzschild Geometry and Generalized Uncertainty Principle'', {\hypersetup{urlcolor=vividviolet}\href{https://link.springer.com/article/10.1140\%2Fepjc\%2Fs10052-019-6610-1}{Eur. Phys. J. C \textbf{79} (2019) 95}}.

\bibitem{1903.01382}
Luca Buoninfante, Giuseppe Gaetano Luciano, Luciano Petruzziello, ``Generalized Uncertainty Principle and Corpuscular Gravity'', \href{https://arxiv.org/abs/1903.01382}{[arXiv:1903.01382 [gr-qc]]}.

\bibitem{1212.0454}
Sabine Hossenfelder, ``Gravity Can be Neither Classical nor Quantized'', In: Aguirre A., Foster B., Merali Z. (eds) Questioning the Foundations of Physics. The Frontiers Collection. Springer, Cham, \href{https://arxiv.org/abs/1212.0454}{[arXiv:1212.0454 [gr-qc]]}.

\bibitem{1208.5874}
Sabine Hossenfelder, ``A Possibility to Solve the Problems with Quantizing Gravity'', {\hypersetup{urlcolor=vividviolet}\href{https://www.sciencedirect.com/science/article/pii/S0370269313006011?via\%3Dihub}{Phys. Lett. B \textbf{725} (2013) 473}}, \href{https://arxiv.org/abs/1208.5874}{[arXiv:1208.5874 [gr-qc]]}.

\bibitem{0610018}
Max Niedermaier, ``The Asymptotic Safety Scenario in Quantum Gravity -- An Introduction'', {\hypersetup{urlcolor=vividviolet}\href{http://iopscience.iop.org/article/10.1088/0264-9381/24/18/R01/meta}{Class. Quant. Grav. \textbf{24} (2007) R171}}, \href{https://arxiv.org/abs/gr-qc/0610018}{[arXiv:gr-qc/0610018]}.

\bibitem{0911.0693}
Brian Greene, Kurt Hinterbichler, Simon Judes, Maulik K. Parikh, ``Smooth Initial Conditions from Weak Gravity'', {\hypersetup{urlcolor=vividviolet}\href{https://www.sciencedirect.com/science/article/pii/S0370269311001250?via\%3Dihub}{Phys. Lett. B \textbf{697} (2011) 178}}, \href{https://arxiv.org/abs/0911.0693}{[arXiv:0911.0693 [hep-th]]}.

\bibitem{9210119}
Gary T. Horowitz, ``The Dark Side of String Theory: Black Holes and Black Strings'', \href{https://arxiv.org/abs/hep-th/9210119}{[arXiv:hep-th/9210119]}.



\bibitem{LV}
P. T. Landsberg, A De Vos, ``The Stefan-Boltzmann Constant in $n$-Dimensional Space'', 	
{\hypersetup{urlcolor=vividviolet}\href{http://iopscience.iop.org/article/10.1088/0305-4470/22/8/021/meta}{Jour. of Phys. A  \textbf{22} (1989) 1073}}.

\bibitem{0510002}
Tatiana R. Cardoso, Antonio S. de Castro, ``The Blackbody Radiation in D-Dimensional Universes'', {\hypersetup{urlcolor=vividviolet}\href{http://www.scielo.br/scielo.php?script=sci_arttext&pid=S1806-11172005000400007&lng=en&tlng=en}{Rev. Bras. Ens. Fis. \textbf{27} (2005) 559}}, \href{https://arxiv.org/abs/quant-ph/0510002}{[arXiv:quant-ph/0510002]}.

\bibitem{0004004}
Roberto Casadio, Benjamin Harms, ``Black Hole Evaporation and Large Extra Dimensions'', {\hypersetup{urlcolor=vividviolet}\href{https://linkinghub.elsevier.com/retrieve/pii/S0370269300008406}{Phys. Lett. B \textbf{487} (2000) 209}}, \href{https://arxiv.org/abs/hep-th/0004004}{[arXiv:hep-th/0004004]}.


\bibitem{9712017}
Roberto Casadio, Benjamn Harms, Yvan Leblanc, ``Microfield Dynamics of Black Holes'', {\changeurlcolor{vividviolet}\href{https://journals.aps.org/prd/abstract/10.1103/PhysRevD.58.044014}{Phys. Rev. D \textbf{58} (1998) 044014}}, \href{https://arxiv.org/abs/gr-qc/9712017}{[arXiv:gr-qc/9712017]}.

\bibitem{1101.1384}
 Roberto Casadio, Benjamin Harms, ``Microcanonical Description of (Micro) Black Holes'', {\changeurlcolor{vividviolet}\href{https://www.mdpi.com/1099-4300/13/2/502}{Entropy \textbf{13} (2011) 502}}, \href{https://arxiv.org/abs/1101.1384}{[arXiv:1101.1384 [hep-th]]}.

\bibitem{0412265}
Sabine Hossenfelder, ``What Black Holes Can Teach Us'', \emph{Focus on Black Hole Research}, pp. 155-192, Nova Science Publishers (2005),
\href{https://arxiv.org/abs/hep-ph/0412265}{[arXiv:hep-ph/0412265]}.

\bibitem{1506.03975}
Finnian Gray, Sebastian Schuster, Alexander Van-Brunt, Matt Visser, ``The Hawking Cascade from a Black Hole Is Extremely Sparse'', {\changeurlcolor{vividviolet}\href{http://iopscience.iop.org/article/10.1088/0264-9381/33/11/115003/meta}{Class. Quant. Grav. \textbf{33} (2016) 115003}}, \href{https://arxiv.org/abs/1506.03975}{[arXiv:1506.03975 [gr-qc]]}.


\bibitem{1512.05809}
Matt Visser, Finnian Gray, Sebastian Schuster, Alexander Van-Brunt, ``Sparsity of the Hawking Flux'', in Proceedings of the MG14 Meeting on General Relativity (2017); pp. 1724-1729, \href{https://arxiv.org/abs/1512.05809}{[arXiv:1512.05809 [gr-qc]]}.

\bibitem{1606.01790}
Wolfgang M\"{u}ck, ``Hawking Radiation is Corpuscular'', {\changeurlcolor{vividviolet}\href{https://dx.doi.org/10.1140/epjc/s10052-016-4233-3}{Eur. Phys. J. C \textbf{76} (2016) 374}}, \href{https://arxiv.org/abs/1606.01790}{[arXiv:1606.01790 [hep-th]]}.

\bibitem{dRGT0} Claudia de Rham, Gregory Gabadadze, ``Generalization of the Fierz-Pauli Action'', {\hypersetup{urlcolor=vividviolet}\href{https://journals.aps.org/prd/abstract/10.1103/PhysRevD.82.044020}{Phys. Rev. D \textbf{82} (2010) 044020}}, \href{https://arxiv.org/abs/1007.0443}{[arXiv:1007.0443 [hep-th]]}.

\bibitem{dRGT} Claudia de Rham, Gregory Gabadadze, Andrew J. Tolley, ``Resummation of Massive Gravity'', {\hypersetup{urlcolor=vividviolet}\href{https://journals.aps.org/prl/abstract/10.1103/PhysRevLett.106.231101}{Phys. Rev. Lett. \textbf{106} (2011) 231101}}, \href{https://arxiv.org/abs/1011.1232}{[arXiv:1011.1232 [hep-th]]}.

\bibitem{Hassan} S. F. Hassan, Rachel A. Rosen, ``On Non-Linear Actions for Massive Gravity'', {\hypersetup{urlcolor=vividviolet}\href{https://link.springer.com/article/10.1007\%2FJHEP07\%282011\%29009}{JHEP \textbf{07} (2011) 009}}, \href{https://arxiv.org/abs/1103.6055}{[arXiv:1103.6055 [hep-th]]}.

\bibitem{1401.4173} Claudia de Rham, ``Massive Gravity'', {\hypersetup{urlcolor=vividviolet}\href{https://link.springer.com/article/10.12942/lrr-2014-7}{Living Rev. Relativ. \textbf{17} (2014) 7}}, \href{https://arxiv.org/abs/1401.4173}{[arXiv:1401.4173 [hep-th]]}.

\bibitem{1610.01505}
S. H. Hendi, N. Riazi, S. Panahiyan, ``Holographical Aspects of Dyonic Black Holes: Massive Gravity Generalization'', {\hypersetup{urlcolor=vividviolet}\href{https://onlinelibrary.wiley.com/doi/abs/10.1002/andp.201700211}{Ann. Phys. (Berlin) \textbf{530} (2018) 1700211}}, \href{https://arxiv.org/abs/1610.01505}{[arXiv:1610.01505 [hep-th]]}.

\bibitem{1510.03204}
Hongsheng Zhang, Xin-Zhou Li,
``Ghost Free Massive Gravity with Singular Reference Metrics'', {\hypersetup{urlcolor=vividviolet}\href{https://journals.aps.org/prd/abstract/10.1103/PhysRevD.93.124039}{Phys. Rev. D 93 (2016) 124039}}, \href{https://arxiv.org/abs/1510.03204}{[arXiv:1510.03204 [gr-qc]]}.

\bibitem{Hawking}
Stephen W. Hawking, ``Black Holes in General Relativity'', {\hypersetup{urlcolor=vividviolet}\href{https://link.springer.com/article/10.1007/BF01877517}{Commun. Math. Phys. \textbf{25} (1972) 152}}.



\bibitem{1306.5457}
Stanley Deser, Keisuke Izumi, Yen Chin Ong, Andrew Waldron, ``Massive Gravity Acausality Redux'', {\hypersetup{urlcolor=vividviolet}\href{https://www.sciencedirect.com/science/article/pii/S0370269313007181?via\%3Dihub}{Phys. Lett. B \textbf{726} (2013) 544}}, \href{https://arxiv.org/abs/1306.5457}{[arXiv:1306.5457 [hep-th]]}.


\bibitem{1410.2289}
Stanley Deser, Keisuke Izumi, Yen Chin Ong, Andrew Waldron, ``Problems of Massive Gravities'', {\hypersetup{urlcolor=vividviolet}\href{https://www.worldscientific.com/doi/abs/10.1142/S0217732315400064}{Mod. Phys. Lett. A \textbf{30} (2015) 1540006}}, \href{https://arxiv.org/abs/1410.2289}{[arXiv:1410.2289 [hep-th]]}.


\bibitem{1505.03518}
Pavel Motloch, Wayne Hu, Austin Joyce, Hayato Motohashi, ``Self-Accelerating Massive Gravity: Superluminality, Cauchy Surfaces and Strong Coupling'', {\hypersetup{urlcolor=vividviolet}\href{https://arxiv.org/abs/1505.03518}{Phys. Rev. D \textbf{92} (2015) 044024}}, \href{https://arxiv.org/abs/1505.03518}{[arXiv:1505.03518 [hep-th]]}.

\bibitem{1603.03423}
Pavel Motloch, Wayne Hu, Hayato Motohashi, ``Self-accelerating Massive Gravity: Hidden Constraints and Characteristics'', 	{\hypersetup{urlcolor=vividviolet}\href{https://journals.aps.org/prd/abstract/10.1103/PhysRevD.93.104026}{Phys. Rev. D \textbf{93} (2016) 104026}}, \href{https://arxiv.org/abs/1603.03423}{[arXiv:1603.03423 [hep-th]]}.

\bibitem{1808.07829}
Behzad Eslam Panah, Seyed Hossein Hendi, Yen Chin Ong, ``Black Hole Remnant in Massive Gravity'', \href{https://arxiv.org/abs/1808.07829}{[arXiv:1808.07829 [gr-qc]]}.

\bibitem{9803061}
Dietmar Klemm, Luciano Vanzo, ``Quantum Properties of Topological Black Holes'', 	{\hypersetup{urlcolor=vividviolet}\href{https://journals.aps.org/prd/abstract/10.1103/PhysRevD.58.104025}{Phys. Rev. D \textbf{58} (1998) 104025}}, \href{https://arxiv.org/abs/gr-qc/9803061}{[arXiv:gr-qc/9803061]}.

\bibitem{1704.02967}
Pablo Bueno, Pablo A. Cano, ``Universal Black Hole Stability in Four Dimensions'', {\hypersetup{urlcolor=vividviolet}\href{https://journals.aps.org/prd/abstract/10.1103/PhysRevD.96.024034}{Phys. Rev. D \textbf{96} (2017) 024034}}, \href{https://arxiv.org/abs/1704.02967}{[arXiv:1704.02967 [hep-th]]}.


\bibitem{1007.1317}
Yi-Fu Cai, Damien A. Easson, ``Black Holes in an Asymptotically Safe Gravity Theory with Higher Derivatives'', {\hypersetup{urlcolor=vividviolet}\href{http://iopscience.iop.org/article/10.1088/1475-7516/2010/09/002/meta}{JCAP \textbf{1009} (2010) 002}}, \href{https://arxiv.org/abs/1007.1317}{[arXiv:1007.1317 [hep-th]]}.



\bibitem{1811.07309v1}
Hao Xu, Man-Hong Yung, ``Black Hole Evaporation in Conformal (Weyl) Gravity'', \href{https://arxiv.org/abs/1811.07309v1}{[arXiv:1811.07309 [gr-qc]]}.

\bibitem{HW}
William A. Hiscock, Lance D. Weems, ``Evolution of Charged Evaporating Black Holes'', {\hypersetup{urlcolor=vividviolet}\href{https://journals.aps.org/prd/abstract/10.1103/PhysRevD.41.1142}{Phys. Rev. D \textbf{41} (1990) 1142}}.

\bibitem{ghs1}
David Garfinkle, Gary T. Horowitz, Andrew Strominger, ``Charged Black Holes in String Theory'',
{\hypersetup{urlcolor=vividviolet}\href{https://journals.aps.org/prd/abstract/10.1103/PhysRevD.43.3140}{Phys. Rev. D \textbf{43} (1991) 31403143}}.

\bibitem{ghs2}
Gary W. Gibbons, ``Antigravitating Black Hole Solutions with Scalar Hair in N = 4 Supergravity'', {\hypersetup{urlcolor=vividviolet}\href{https://www.sciencedirect.com/science/article/pii/0550321382901705}{Nucl. Phys. B \textbf{207} (1982) 337}}.

\bibitem{ghs3}
Gary W. Gibbons, Kei-ichi Maeda, ``Black Holes and Membranes in Higher Dimensional Theories with Dilaton Fields'', {\hypersetup{urlcolor=vividviolet}\href{https://www.sciencedirect.com/science/article/pii/0550321388900065?via\%3Dihub}{Nucl. Phys. B \textbf{298} (1988) 741}}.

\bibitem{Gibbons}
Gary W. Gibbons, ``Vacuum Polarization and the Spontaneous Loss of Charge by Black Holes'', {\hypersetup{urlcolor=vividviolet}\href{https://link.springer.com/article/10.1007/BF01609829}{Comm. Math. Phys. \textbf{44} (1975) 245}}.



\bibitem{0804.0055}
Jorge V. Rocha, ``Evaporation of Large Black Holes in AdS: Coupling to the Evaporon'', {\hypersetup{urlcolor=vividviolet}\href{http://iopscience.iop.org/article/10.1088/1126-6708/2008/08/075/meta}{JHEP \textbf{08} (2008) 075}}, \href{https://arxiv.org/abs/0804.0055}{[arXiv:0804.0055 [hep-th]]}.

\bibitem{1304.6483}
Ahmed Almheiri, Donald Marolf, Joseph Polchinski, Douglas Stanford, James Sully, ``An Apologia for Firewalls'', {\hypersetup{urlcolor=vividviolet}\href{https://link.springer.com/article/10.1007\%2FJHEP09\%282013\%29018}{JHEP \textbf{09} (2013) 018}}, \href{https://arxiv.org/abs/1304.6483}{[arXiv:1304.6483 [hep-th]]}.

\bibitem{1307.1796}
Mark Van Raamsdonk, ``Evaporating Firewalls'', {\hypersetup{urlcolor=vividviolet}\href{https://link.springer.com/article/10.1007\%2FJHEP11\%282014\%29038}{JHEP \textbf{11} (2014) 038}}, \href{https://arxiv.org/abs/1307.1796}{[arXiv:1307.1796 [hep-th]]}.

\bibitem{1507.02682}
Don N. Page, ``Finite Upper Bound for the Hawking Decay Time of an Arbitrarily Large Black Hole in Anti-de Sitter Spacetime'', {\hypersetup{urlcolor=vividviolet}\href{https://journals.aps.org/prd/abstract/10.1103/PhysRevD.97.024004}{Phys. Rev. D \textbf{97} (2018) 024004}}, \href{https://arxiv.org/abs/1507.02682}{[arXiv:1507.02682 [hep-th]]}.

\bibitem{1507.07845}
Yen Chin Ong, ``Hawking Evaporation Time Scale of Topological Black Holes in Anti-de Sitter Spacetime'', {\hypersetup{urlcolor=vividviolet}\href{https://www.sciencedirect.com/science/article/pii/S0550321316000067?via\%3Dihub}{Nucl. Phys. B \textbf{903} (2016) 387}}, \href{https://arxiv.org/abs/1507.07845}{[arXiv:1507.07845 [gr-qc]]}.

\bibitem{9808032}
Danny Birmingham, ``Topological Black Holes in Anti-de Sitter Space'', {\hypersetup{urlcolor=vividviolet}\href{http://iopscience.iop.org/article/10.1088/0264-9381/16/4/009/meta}{Class. Quant. Grav. \textbf{16} (1999) 1197}}, \href{https://arxiv.org/abs/hep-th/9808032}{[arXiv:hep-th/9808032]}.


\bibitem{bch}
James M. Bardeen, Brandon Carter, Stephen W. Hawking, ``The Four Laws of Black Hole Mechanics'', {\hypersetup{urlcolor=vividviolet}\href{https://projecteuclid.org/euclid.cmp/1103858973}{Comm. Math. Phys. \textbf{31}  (1973) 161}}.


\bibitem{1411.2854}
Marios Christodoulou, Carlo Rovelli, ``How Big is a Black Hole?'', {\hypersetup{urlcolor=vividviolet}\href{https://dx.doi.org/10.1103/PhysRevD.91.064046}{Phys. Rev. D \textbf{91} (2015) 064046}}, \href{https://arxiv.org/abs/1411.2854}{[arXiv:1411.2854 [gr-qc]]}.

\bibitem{1503.01092}
Yen Chin Ong, ``Never Judge a Black Hole by Its Area'', {\hypersetup{urlcolor=vividviolet}\href{http://iopscience.iop.org/article/10.1088/1475-7516/2015/04/003/meta}{JCAP \textbf{04} (2015) 003}}, \href{https://arxiv.org/abs/1503.01092}{[arXiv:1503.01092 [gr-qc]]}.

\bibitem{9405007v2}
Renaud Parentani, Tsvi Piran, ``The Internal Geometry of an Evaporating Black Hole'', {\hypersetup{urlcolor=vividviolet}\href{https://journals.aps.org/prl/abstract/10.1103/PhysRevLett.73.2805}{Phys. Rev. Lett. \textbf{73} (1994) 2805}}

\bibitem{1503.08245}
Yen Chin Ong, ``The Persistence of the Large Volumes in Black Holes'', {\hypersetup{urlcolor=vividviolet}\href{https://link.springer.com/article/10.1007\%2Fs10714-015-1929-x}{Gen. Relativ. Gravit. \textbf{47} (2015) 88}}, \href{https://arxiv.org/abs/1503.08245}{[arXiv:1503.08245 [gr-qc]]}.

\bibitem{1604.07222}
Marios Christodoulou, Tommaso De Lorenzo, ``On the Volume Inside Old Black Holes'', {\hypersetup{urlcolor=vividviolet}\href{https://link.aps.org/doi/10.1103/PhysRevD.94.104002}{Phys. Rev. D \textbf{94} (2016) 104002}}, \href{https://arxiv.org/abs/1604.07222}{[arXiv:1604.07222 [gr-qc]]}.


\bibitem{davies}
Paul C. W. Davies, ``Thermodynamics of Black Holes'',  {\hypersetup{urlcolor=vividviolet}\href{http://iopscience.iop.org/article/10.1088/0034-4885/41/8/004/pdf}{Rep. Prog. Phys. \textbf{41} (1978) 1313}}.

\bibitem{FH}
Charles J. Farrugia, Petr H\'{a}j\'{i}\v{c}ek , ``The Third Law of Black Hole Mechanics: A Counterexample'', {\hypersetup{urlcolor=vividviolet}\href{https://link.springer.com/article/10.1007/BF01221129}{Commun. Math. Phys. \textbf{68} (1979) 291}}.

\bibitem{P}
Mieczysław Pr\'oszy\'nsk, ``Thin Charged Shells and the Violation of the Third Law of Black Hole Mechanics'', {\hypersetup{urlcolor=vividviolet}\href{https://link.springer.com/article/10.1007/BF0075993}{Gen Relat. Gravit. \textbf{15} (1983) 403}}.

\bibitem{Torii}
Takashi Torii, ``Violation of the Third Law of Black Hole Thermodynamics in Higher Curvature Gravity'', {\hypersetup{urlcolor=vividviolet}\href{https://www.mdpi.com/1099-4300/14/11/2291/htm}{Entropy \textbf{14} (2012) 229}}.


\end{thebibliography}
\end{document}